\documentclass[twocolumn,tightenlines,prd,aps,superscriptaddress]{revtex4}
\usepackage{epsf}

\begin{document}

\preprint{LA-UR-02-5306}
\title{The Inflationary Perturbation Spectrum}

\author{Salman Habib}
\author{Katrin Heitmann}
\author{Gerard Jungman}

\affiliation{Theoretical Division, MS B285, Los Alamos National
Laboratory, Los Alamos, New Mexico 87545}

\author{Carmen Molina-Par\'{\i}s}
\affiliation{Mathematics Institute, University of Warwick,
Coventry CV4 7AL, U.K.}

\date{\today}

\begin{abstract}
Motivated by the prospect of testing inflation from precision cosmic
microwave background observations, we present analytic results for
scalar and tensor perturbations in single-field inflation models based
on the application of uniform approximations. This technique is
systematically improvable, possesses controlled error bounds, and does
not rely on assuming the slow-roll parameters to be constant. We
provide closed-form expressions for the power spectra and the
corresponding scalar and tensor spectral indices.
\end{abstract}


\maketitle

\vskip2pc

Recent observations of the cosmic microwave background (CMB) and of
the distribution of galaxies have been largely consistent with
inflation, not only with the result that the Universe is at critical
density but also with the idea that the structure in the Universe
arose from the gravitational collapse of adiabatic, Gaussian, and
nearly-scale invariant primordial fluctuations, a key prediction of
simple models of inflation \cite{infpertorig}. The prospect of using
precision CMB observations to distinguish between models of inflation
has focused attention on the predictive control of the theory
underlying the generation of inflationary fluctuations.

The form of the power spectrum alluded to above is predicted by models
where inflation is caused by the dynamics of a scalar field in a
`featureless' potential, evolving in a friction-dominated `slow-roll'
regime. In more complex models (multiple fields, nontrivial
potentials) it is possible to engineer spectra in a variety of ways;
however, present observations do not demand the consideration of
dynamically exotic inflation models.

Inflation predicts a spectrum of metric perturbations in the scalar
(density) and tensor (gravitational wave) sectors, the vector
component being naturally suppressed. Both scalar and tensor
perturbations cause anisotropy in the CMB temperature
\cite{anisotropy}. Tensor modes cause a potentially observable
polarization of the CMB, a prime target for next-generation CMB
observations \cite{polar}.

In tests of the inflationary paradigm, a key point is that, in
inflation, the scalar and tensor fluctuations are not independent. It
is important to accurately quantify this relationship, e.g., via
`consistency relations' \cite {cons}. However, approximations based on
`slow-roll' assumptions lack error control and are not systematically
improvable \cite{wms}. Attempts to cure this problem have so far led
to restrictive assumptions and/or tedious mathematical formulations.

Our purpose in this Letter is to present an analysis of the
inflationary perturbation spectrum by applying a uniform approximation
to the relevant mode equations. Our analysis does not make restrictive
assumptions, possesses error bounds, and is systematically
improvable. We also provide compact expressions for the scalar and
tensor spectral indices $n_S$ and $n_T$. Details will be reported
elsewhere \cite{hhjmp}.

The dynamical equations for gauge-invariant scalar perturbations can
be found by linearizing the Einstein equations in the longitudinal
gauge (see, e.g., Ref.~\cite{ivrevs2}). In terms of the standard
choice of gauge invariant variable $u$ (following the conventions of
Ref.~\cite{sl}), one of these equations reduces to the compact form
$u^{\prime\prime}-\Delta u-{(z^{\prime\prime}/z)}u=0$ where $z\equiv
a\phi^{\prime}/h$, $h\equiv a^{\prime}/a$, $a$ is the scale factor and
the prime denotes a derivative with respect to the conformal time
$\eta\equiv \int dt/a$. This equation represents the dynamics of a
free field with a time-dependent mass. The canonical form makes $u$
the appropriate variable to quantize.

To proceed further, we introduce the quantity $\zeta\equiv u/z$, the
intrinsic curvature perturbation of the comoving hypersurfaces
\cite{givdhl}. The dynamical equation for $u$ forces $\zeta$ to be
nearly constant in the long wavelength limit $k\rightarrow 0$. This is
true during the inflationary phase as well as in the post-reheating
era. A computation of $\zeta$ provides all the information needed
(aside from the transfer functions) to extract the temperature
anisotropy of the CMB. Details of this procedure can be found in
Refs. \cite{ivrevs2,wms,ms2}.

The calculation of power spectra involves computing two-point
functions of quantum operators, e.g.,
\begin{equation}
\langle 0|\hat{u}(\eta,{\bf x})\hat{u}(\eta,{\bf x}+{\bf
r})|0\rangle=\int_0^{\infty} {dk\over k} {\sin kr\over kr}
P_u(\eta,k).
\label{pspectra}
\end{equation}
The associated complex amplitude $u_k(\eta)$ satisfies
\begin{equation}
u_k^{\prime\prime}+\left(k^2 -{z^{\prime\prime}\over z}\right)u_k=0~.
\label{mode}
\end{equation}
Solving Eqn. (\ref{mode}) is the fundamental problem in determining
the primordial power spectrum $P_u$ (or the power spectrum for
$\zeta$, here defined as $P_S$).

The case of the tensor perturbations proceeds analogously. The
relevant mode equations are
\begin{equation}
v^{\prime\prime}_{k} + \left(k^2 - {a^{\prime\prime} \over
a}\right) v_{k}=0~,
\label{givgw}
\end{equation}
which correspond to the wave equation for a massless, minimally
coupled, scalar field in an FRW Universe.

One approach to analytically approximating Eqns. (\ref{mode}) and
(\ref{givgw}) relies on the fact that exact solutions exist in the
limits $k^2 \gg \left|z^{\prime\prime}/z\right|$ (short wavelength)
and $k^2\ll \left|z^{\prime\prime}/z\right|$ (long wavelength) or, as
will be made more explicit below, as $-k\eta\rightarrow \infty$ and
$k\eta\rightarrow 0^-$. For scalar perturbations,
\begin{eqnarray}
u_k&\rightarrow&{{\hbox{e}}^{-ik\eta}\over\sqrt{2k}}~~\left(k^2 \gg
\left|z^{\prime\prime}/z\right|,~-k\eta\rightarrow \infty\right),
\label{asymp1}\\
u_k&\rightarrow&A_k
z~~\left(k^2\ll\left|z^{\prime\prime}/z\right|,~k\eta\rightarrow
0^-\right). 
\label{asymp2}
\end{eqnarray}
Here, the short wavelength solution corresponds to the choice of an
adiabatic vacuum for modes on length scales much smaller than the
scale set by the curvature. The long wavelength solutions correspond
to the growing mode on scales much larger than the Hubble
length. Similar solutions exist for tensor perturbations. The problem
of determining the power spectrum boils down to matching the
asymptotic solutions, i.e., determining $A_k$.

To facilitate further analysis, it is useful to introduce the
substitution 
\begin{equation}
{z^{\prime\prime}/z}\equiv \left(\nu_S^2(\eta)-{1/4}\right)/\eta^2
\label{nudef}
\end{equation}
in Eqn. (\ref{mode}), yielding
\begin{equation}
u_k^{\prime\prime}(\eta)+\left[k^2-{1\over
\eta^2}\left(\nu_S^2(\eta)-{1\over 4}\right)\right]u_k(\eta)=0.
\label{modeua}
\end{equation}
In the standard slow-roll method, $\nu_S(\eta)$ is taken to be
constant in the intermediate regime $-k\eta\sim 1$ or $k\sim aH$
(horizon-crossing for the $k$-mode of interest), in which case the
mode equations can be solved in terms of Bessel functions and these
intermediate solutions matched to the short and long wavelength
solutions \cite{sl} (see the discussion in Ref.~\cite{wms}).
Systematic improvement of the approach runs into serious difficulties
\cite{wms}: as soon as $\nu_S(\eta)$ is varying, the Bessel solution
no longer holds. (A perturbative improvement based on the leading
order Bessel solution is given in Ref.~\cite{sg}. See also,
Ref. {\cite{seqns}.)

To overcome these difficulties, we will here apply the method of
uniform approximation \cite{Olver}. This method provides a single
approximating solution which converges uniformly, for time-dependent
$\nu_S$, taking on the Liouville-Green (LG) or WKB form in both the
short and long wavelength limits. The uniform approximation employs no
intermediate matching, and, as opposed to the standard slow-roll
approach, the approximation procedure can be systematically improved
and possesses an error control function~\cite{hhjmp}.

We now turn to the specific form of the approximate
solutions. Following Ref. \cite{Olver} we find a solution for
$u_k(\eta)$ containing a part valid to the left of the turning point,
$\eta=\bar{\eta}_S$ (defined by $k^2=\nu_S^2/\bar{\eta}_S^2$), and a
part valid to the right. The leading order unnormalized solutions to
the left $(\eta\leq\bar\eta_S)$ and right $(\eta\geq\bar\eta_S)$ of
the turning point can be written in terms of Airy functions (${\rm
Ai^{(1)}}\equiv {\rm Ai}$, ${\rm Ai^{(2)}}\equiv {\rm Bi}$) as
\begin{eqnarray}
u^{(1,2)}_{\stackrel{k<}{>}}(\eta)&=&f_{\stackrel{<}{>}}^{1/4}(\eta)
g_S^{-1/4}(\eta){\rm Ai^{(1,2)}}[f_{\stackrel{<}{>}}(\eta)],
\label{uniformapp}\\
f_{\stackrel{<}{>}}(\eta)&\equiv&\mp\left\{\pm\frac 3
2\int_\eta^{\bar\eta_S}d \eta'\left[\mp 
g_S(\eta^{\prime})\right]^{1/2}\right\}^{2/3},\\
g_S(\eta)&\equiv&\frac{\nu_S^2(\eta)}{\eta^2}-k^2.
\end{eqnarray}
The general solution is a linear combination $u_k=Au_k^{(1)} +
Bu_k^{(2)}$ with the constants $A$ and $B$ ensuring that
$u_k(\eta)=(1/\sqrt{2k})e^{-ik\eta}$ in the limit
$k\rightarrow\infty$. In this limit, for well-behaved $\nu_S$,
$f_<(\eta)$ is large and negative and we can employ the asymptotic
forms
\begin{equation}
\{{\rm Ai^{(1)}},{\rm Ai^{(2)}}\}(-x) =
{\{\cos,\sin\}\left(2x^{3/2}/3-\pi/4\right)\over\pi^{1/2}x^{1/4}}.
\label{kbig}
\end{equation}
Making the choices, $A=\sqrt{\pi/2}$, $B=-i\sqrt{\pi/2}$, we obtain
the required adiabatic form of the solution at short wavelengths and
as $\eta\rightarrow-\infty$ \cite{fulling}. Note how the uniform
approximation makes the matching procedure very simple.

\begin{figure}[hb]
\begin{center}
\parbox{7.5 cm}{
\mbox{\mbox{\epsfxsize=7.5 cm\epsfbox{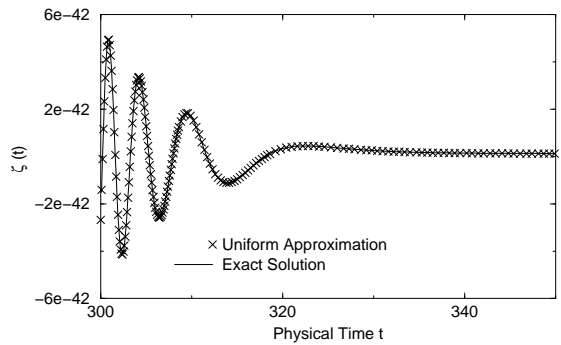}}}}
\caption{Comparison of $\zeta$ for a $\phi^2$ chaotic inflation model
calculated using the uniform approximation (crosses) and a numerical
solution of the exact equations (solid line).} 
\label{fig1}
\end{center}
\end{figure}

The power spectra are computed in the limit $\eta\rightarrow
0^-$. The $1/\eta^2$ pole dominates the behavior of the solutions
and the Airy solution limits to the LG/WKB solution.
The argument of the Airy functions, $f_>(\eta)$, becomes large,
allowing the use of the asymptotic forms
\begin{equation}
\{{\rm Ai^{(1)}},{\rm Ai^{(2)}}\}(x)=
{x^{-1/4}\over{\{2,1\}\sqrt{\pi}}}~\exp\left(\{-,+\}\frac{2}{3}
x^{3/2}\right).
\end{equation}
For computing the power spectra in the $k\eta\rightarrow 0^-$ limit
only the growing part of the solution is relevant:
\begin{equation}
u_k(\eta)\stackrel{k\eta\rightarrow 0^-}{=}-i{C}
\sqrt{-{\eta\over\nu_S(\eta)}}\exp\left(\int_{\bar\eta_S}^\eta
d\eta^{\prime}\sqrt{g_S(\eta^{\prime})}\right),
\label{lguk}
\end{equation}
where $C$ is an irrelevant constant phase factor. The calculation for
the tensor case proceeds in a similar fashion, with the replacement of
$\nu_S$ by $\nu_T$ defined by $a^{\prime\prime}/a\equiv
\left[\nu_T^2(\eta)-1/4\right]/\eta^2$. A numerical illustration of the
accuracy of the approximation is provided in Fig. \ref{fig1}.

Once the approximate solutions to Eqns. (\ref{mode}) and (\ref{givgw})
have been found in the manner described above, the relevant power
spectra can be easily computed via $P_S(k)\equiv
(k^3/2\pi^2)|u_k(\eta)/z|^2$. Using Eqn. (\ref{lguk}), we find
\begin{equation}\label{powersp}
P_S(k)=\frac{k^3}{4\pi^2}\frac{1}{|z(\eta)|^2}
\frac{-\eta}{\nu_S(\eta)} \exp\left(2\int_{\bar\eta_S}^\eta
d\eta^{\prime}\sqrt{g_S(\eta^{\prime})}\right).\label{psps}
\end{equation}
The generalized spectral index for scalar perturbations can be
obtained from the power spectrum via $n_S(k)=1+{d\ln P_S(k)/d\ln k}$.
It is important to note that the turning point is a function of $k$,
since $k=-\nu_S(\bar\eta_S)/\bar\eta_S$ where
$\nu_S(\bar\eta_S)$ is the value of $\nu_S(\eta)$ at the turning
point $\eta=\bar\eta_S$. Using this relation, one finds
\begin{equation}
n_S(k)=4+2\frac{\nu_S(\bar\eta_S) k}{|\bar\eta_S|}
\int_{\bar\eta_S}^\eta\frac{d\eta'}
{\sqrt{g_S(\eta^{\prime})}}.
\label{nsint}
\end{equation}
The above analysis can be carried out for tensor perturbations in an
identical fashion with the replacement $\nu_S\rightarrow\nu_T$. The
spectral index for tensors is given by
\begin{equation}
n_T(k)=3+2\frac{\nu_T(\bar\eta_T)k}{|\bar\eta_T|}
\int_{\bar\eta_T}^\eta\frac{d\eta'}
{\sqrt{g_T(\eta^{\prime})}}.
\label{ngint}
\end{equation}
The expressions (\ref{psps}), (\ref{nsint}), and (\ref{ngint})
constitute our fundamental result.

In order to achieve a completely local expression, we can make
additional approximations in the integrals in Eqns. (\ref{nsint}) and
(\ref{ngint}). The integrand has a square-root singularity at the
turning point while at the upper limit $\eta$ goes to zero and the
integrand vanishes linearly. Therefore, assuming $\nu_S(\eta)$ and
$\nu_T(\eta)$ are well-behaved, we expect the dominant contribution to
arise from the lower limit. Assuming that $\nu_S$ and $\nu_T$ vary in
such a way that one can expand them around the turning point in a
Taylor series, to first order in derivatives, we find
\begin{equation}
\nu_S^2(\eta)\simeq
\nu_S^2(\bar\eta_S)+2\nu_S(\bar\eta_S)
\nu'_S(\bar\eta_S)\left(\eta-\bar\eta_S\right).
\end{equation}
The integrals in Eqns. (\ref{nsint}) and (\ref{ngint}) can now be
solved exactly, and the spectral indices are given by
\begin{eqnarray}
n_S(k)&=&4-2\nu_S(\bar\eta_S)\left[1-\left(1-\frac\pi 2\right)
\frac{\nu'_S(\bar\eta_S)}{\nu_S(\bar\eta_S)}\bar\eta_S\right],
\label{nslocal}\\
n_T(k)&=&3-2\nu_T(\bar\eta_T)\left[1-\left(1-\frac\pi 2\right)
\frac{\nu'_T(\bar\eta_T)}{\nu_T(\bar\eta_T)}\bar\eta_T\right].
\label{ntlocal}
\end{eqnarray}

As a simple test of the uniform approximation, we consider the case of
constant $\nu_S$ and $\nu_T$ where exact results are available. In
this case (\ref{psps}), (\ref{nsint}), (\ref{ngint}) can be evaluated
immediately. One finds
\begin{eqnarray}
P_S(k)&=&\frac{2^{2\nu_S-2}}{\pi^2}\nu_S^{2\nu_S-1}e^{-2\nu_S}
\left(\frac{H}{a\dot\phi}\right)^2
(-k\eta)^{-2\nu_S+1} k^2\nonumber\\
&=&P_S^{\rm ex}(k)\left(1-\frac{1}{6\nu_S}+
\frac{1}{72\nu^2_S}+\cdots\right),\label{PSster}\\
n_S&=&4-2\nu_S,~~~~~n_T=3-2\nu_T,
\end{eqnarray}
where we have used Stirling's formula for the Gamma function in the
exact power spectrum $P_S^{\rm ex}(k)$ (See, e.g.,
Ref.~\cite{recon}). The spectral indices are found exactly, unlike in
slow-roll expansions. All analytic examples so far used to test
slow-roll results \cite{sl,sg}, e.g., power-law inflation, inflation
near a maximum, and natural inflation have constant $\nu_S$: the
uniform approximation recovers the spectral indices exactly in all of
these cases.

Error bounds follow from the computation of the variation of the error
control function~\cite{Olver}
\begin{eqnarray}
{\cal E}(\eta)&=&-{1\over 4}\int d\eta\left\{g_S^{-3/2}\left[g_S''-{5\over
4}g_S^{-1}(g_S')^2-{1\over\eta^2}g_S\right]\right\}\nonumber\\
&&\pm{5\over 24|f_{\stackrel{<}{>}}|^{3/2}}.
\end{eqnarray}
This calculation allows the determination of the errors
$\epsilon_1(\eta)$ and $\epsilon_2(\eta)$ in the two independent
solutions $u_k^{(1)}$ and $u_k^{(2)}$ (\ref{uniformapp}). The error
can be reduced if required by applying higher-order uniform
approximations. Details will be provided in Ref.~\cite{hhjmp}. As an
illustration, for the case of constant $\nu_S$, the absolute value of
the relative error in the unnormalized solutions, over the full domain
of interest, is bounded by
\begin{equation}
|\epsilon_{1,2}|\leq \sqrt{2}\left({1\over 6\nu_S}+{\lambda\over
 72\nu^2_S}+\cdots\right),
\end{equation}
where $\lambda\simeq 1.04$. The error bound has a weak subleading
dependence on $k$ arising from the end-points, falling as $1/k\eta$
far to the left of $\bar\eta_S$ and vanishing as $k^2\eta^2$, in the
limit $\eta\rightarrow 0^-$. The fact that the error bound is
essentially $k$-independent is consistent with the spectral indices
being exact in this case. The error in the power spectrum given in
Eqn.~(\ref{PSster}) falls comfortably within the bound. 

It is instructive to recast our results in the more familiar language
of the slow-roll expansion. The exact forms of $z^{\prime\prime}/z$
and $a^{\prime\prime}/a$ are
\begin{eqnarray}
{z^{\prime\prime}\over z}&=&2a^2H^2\left(1+\epsilon-{3\over 2}\delta
-{1\over 2}\epsilon\delta+{1\over 2}\delta^2+{1\over 2}\xi_2\right),
\label{zdpz} \\
{a^{\prime\prime}\over a}&=&2a^2H^2\left(1-{1\over
2}\epsilon\right), \label{adpa}
\end{eqnarray}
where $\epsilon\equiv-{\dot{H}/H^2}$,
$\delta\equiv-{\ddot{\phi}/(H\dot{\phi})}$, and
$\xi_2\equiv{(\dot{\epsilon}-\dot{\delta})/H}$.
The overdot represents a derivative with respect to the cosmic
time $t$ and $H\equiv\dot{a}/a$.

In the slow-roll approximation $\epsilon$, $\delta$, and $\xi_2$ are
treated as small parameters, and one aims to solve Eqns. (\ref{mode})
and (\ref{givgw}) in terms of expansions in these parameters. The
exact equations of motion for $\epsilon$ and $\delta$ are
\begin{equation}
{\dot{\epsilon}\over
H}=2\epsilon(\epsilon-\delta),~~{\dot{\delta}\over
H}=2\epsilon(\epsilon-\delta)-\xi_2.
\label{srdyn}
\end{equation}
The conformal time $\eta$ can be written in terms of the slow-roll
parameters via repeated integration by parts~\cite{recon}:
\begin{equation}
\eta=-{1\over aH}\left[{1\over
1-\epsilon}+2\epsilon(\epsilon-\delta)+\cdots\right].
\label{etasr}
\end{equation}
If $\epsilon$, $\delta \ll 1$, $\xi_2\sim{\cal{O}}(\epsilon^2,
\delta^2, \epsilon\delta)$, then, $\epsilon$ and $\delta$ are constant
to leading order (and to this order, $\xi_2\simeq 0$; the notation
$\xi_2$ makes explicit that $\xi_2$ is a second-order quantity). At
leading order, $\eta\simeq-(1+\epsilon)/aH$
[Cf. Eqn. ({\ref{etasr}})].

Our results (\ref{nslocal}) and (\ref{ntlocal}) correspond to a
resummation of the slow-roll expansion to all orders. The procedure of
partially undoing this resummation to produce a less-accurate, but
more familiar, expansion is as follows. We begin by substituting
Eqn.~(\ref{zdpz}) in the exact expression for $\nu^2_S$,
Eqn.~(\ref{nudef}).  Expanding Eqn.~(\ref{etasr}) to quadratic order
in slow-roll parameters, $a^2H^2\simeq\
(1+2\epsilon+7\epsilon^2-4\epsilon\delta)/\eta^2$, then yields,
\begin{eqnarray}\label{nusr}
\nu_S&\simeq&\frac 3 2+2\epsilon-\delta-\frac{11}{3}\epsilon\delta
+\frac{14}{3}\epsilon^2+\frac{1}{3}\xi_{2},\nonumber\\
\eta\nu_S'&\simeq& 2\epsilon(\delta-\epsilon)-2\xi_2.
\end{eqnarray}
The desired form of the final result can be obtained by substituting
these expressions in Eqn. (\ref{nslocal})
\begin{eqnarray}
n_S(k)&\simeq &1-4\epsilon_0+2\delta_0
+2\left(\frac{17}{3}-\pi\right)\epsilon_0\delta_0\nonumber\\
&&-2\left(\frac{20}{3}-\pi\right)
\epsilon^2_0-2\left(\frac 4 3-\frac \pi 2\right)\xi_{02},
\end{eqnarray}
where the subscripts on the slow-roll parameters denote that they are
evaluated at the turning point
$k=-\nu_S(\bar\eta_S)/\bar\eta_S$. Clearly this expression is of  
lower accuracy than (\ref{nslocal}). For example, one no longer
recovers the exact results when $\nu_S$ is constant. The analogous
result for the tensor spectral index is
\begin{equation}
n_T(k)\simeq -2\epsilon_0-2\left(\frac{23}{4}-\pi\right)\epsilon_0^2
+2\left(\frac{14}{3}-\pi\right)
\epsilon_0\delta_0.
\end{equation}
These second-order slow-roll results are new and can be compared
against the results of Ref.~\cite{sg}: both agree at leading order as
well as for the case of power-law inflation, however the
next-to-leading expressions differ in the general case. The second
order approximations given above are consistent with the first order
derivative expansions in Eqns.~(\ref{nslocal}) and (\ref{ntlocal}).
The consistency of higher order derivative approximations can be
determined by comparing against the computed error bounds at a given
order of the uniform approximation.

To summarize, we have presented a controlled approximation for the
computation of the inflationary perturbation spectrum which overcomes
the matching problem of the conventional slow-roll approximation and
systematically improves the calculation in the generic case of varying
$\nu_S$ and $\nu_T$. Future applications of this method include
precision studies of generic predictions of inflation
models~\cite{kinney}, reconstruction~\cite{recon}, and studies of
backreaction of quantum fluctuations on the background spacetime
metric.

We thank Scott Dodelson, Fabio Finelli, Sabino Matarrese, Dominik
Schwarz, Mike Turner, and, especially, Ewan Stewart for helpful
conversations. SH and KH acknowledge the hospitality of the Aspen
Center for Physics.

\end{document}